\documentclass[12pt]{iopart}

\usepackage{bm,color,bbm}
\usepackage{graphicx}   
\usepackage{verbatim}   
\usepackage{color}      
\usepackage{subfigure}  
\usepackage{hyperref}   
\usepackage{booktabs}
\usepackage{threeparttable}
\usepackage{dcolumn}
\usepackage{multirow}
\usepackage{array}
\usepackage[ruled]{algorithm2e}

\newtheorem{algo}{Algorithm}

\newcommand{\beq}{\begin{equation}}
\newcommand{\eeq}{\end{equation}}
\newcommand{\bqa}{\begin{eqnarray}}
\newcommand{\eqa}{\end{eqnarray}}

\definecolor{maroon}{rgb}{0.7,0,0}
\definecolor{ngreen}{rgb}{0.3,0.7,0.3}

\definecolor{golden}{rgb}{0.8,0.6,0.1}

\graphicspath{ {./figures/} }

\begin{document}

\title[]{Detecting quantum entanglement with unsupervised learning}

\author{Yiwei Chen$^1$, Yu Pan$^1$$^,$$^*$, Guofeng Zhang$^2$$^,$$^3$$^,$$^*$, Shuming Cheng$^4$$^,$$^5$$^,$$^6$$^,$$^*$}

\address{$^1$Institute of Cyber-Systems and Control, College of Control Science and Engineering, Zhejiang University, Hangzhou, 310027, P. R. China}
\address{$^2$Department of Applied Mathematics, The Hong Kong Polytechnic University, Hung Hom, Kowloon, Hong Kong}
\address{$^3$The Hong Kong Polytechnic University Shenzhen Research Institute, Shenzhen 518057, P. R. China}
\address{$^4$The Department of Control Science and Engineering, Tongji University, Shanghai 201804, P. R. China}
\address{$^5$Shanghai Institute of Intelligent Science and Technology, Tongji University, Shanghai 201804, P. R. China}
\address{$^6$Institute for Advanced Study, Tongji University, Shanghai, 200092, P. R. China}
\address{$^*$Authors to whom any correspondence should be addressed}

\eads{\mailto{ypan@zju.edu.cn}}
\eads{\mailto{Guofeng.Zhang@polyu.edu.hk}}
\eads{\mailto{shuming\_cheng@tongji.edu.cn}}
\vspace{10pt}
\begin{indented}
\item[]September 2021
\end{indented}

\begin{abstract}
Quantum properties, such as entanglement and coherence, are indispensable resources in various quantum information processing tasks. However, there still lacks an efficient and scalable way to detecting these useful features especially for high-dimensional and multipartite quantum systems. In this work, we exploit the convexity of samples without the desired quantum features and design an unsupervised machine learning method to detect the presence of such features as anomalies. Particularly, in the context of entanglement detection, we propose a complex-valued neural network composed of pseudo-siamese network and generative adversarial net, and then train it with only separable states to construct non-linear witnesses for entanglement. It is shown via numerical examples, ranging from 2-qubit to 10-qubit systems, that our network is able to achieve high detection accuracy which is above $97.5\%$ on average. Moreover, it is capable of revealing rich structures of entanglement, such as partial entanglement among subsystems. Our results are readily applicable to the detection of other quantum resources such as Bell nonlocality and steerability, and thus our work could provide a powerful tool to extract quantum features hidden in multipartite quantum data.
\end{abstract}

\section{Introduction}

\begin{figure}[t]
	\includegraphics[scale=0.6]{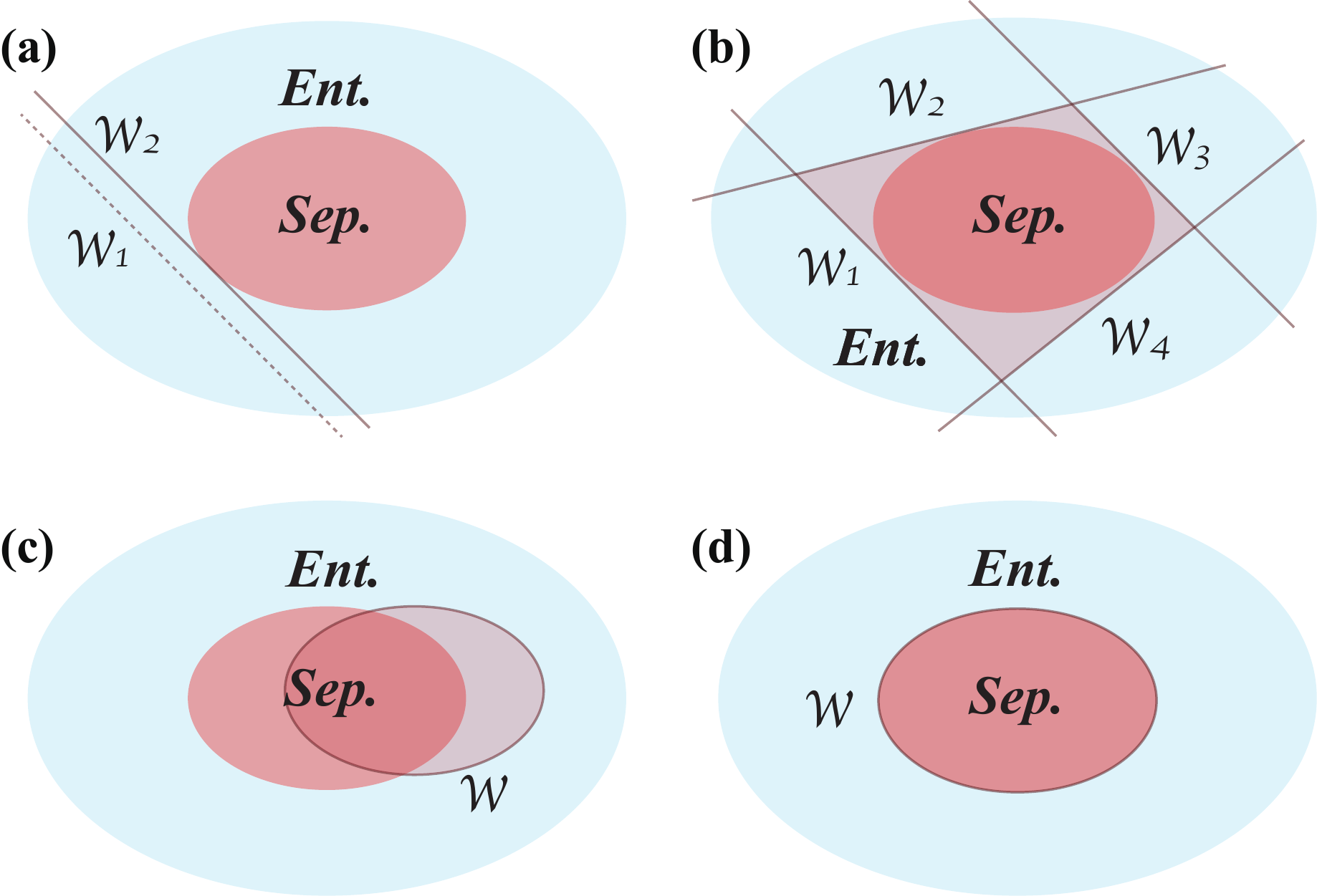}
	\centering
	\caption{
		\label{Fig:entanglement-witness}
		\textbf{The geometrical interpretation of entanglement detection via entanglement witnesses.}
		\textbf{(a)} Standard linear entanglement witnesses. The witness $\mathcal{W}_2$ is finer than $\mathcal{W}_1$.
		\textbf{(b)} A proper set of linear entanglement witnesses is able to form a closed area which encloses all separable states. Here $\mathcal{W}_1, \dots,\mathcal{W}_4$ are used as examples.
		\textbf{(c)} An imperfect nonlinear entanglement witness could be generated via supervised learning method if we cannot label enough samples to cover the space of entangled states.
		\textbf{(d)} A near-perfect nonlinear witness $\mathcal{W}$ can be approximately constructed by the unsupervised neural network if the generated training data span the space of separable states.
	}
\end{figure}

\noindent
Peculiar quantum features, signalled by quantum entanglement~\cite{horodecki2009quantum} and coherence~\cite{Streltsov2017}, enable us to accomplish tasks impossible for classical systems~\cite{Chitambar2019}, such as ensuring the security of communications and speeding up certain hard computational tasks~\cite{Nielson2000, deutsch2020harnessing}. Hence, an important question naturally arises: How can the presence of these features be efficiently detected for any given quantum system? Indeed, this is a challenging task for high-dimensional and multipartite systems because quantum features usually imply correlated patterns hidden within subsystems. Taking entanglement for example, except for low-dimensional systems, e.g., $2 \otimes 2$ and $2 \otimes 3$, of which entanglement could be detected faithfully via the Positive Partial Transpose (PPT) criterion~\cite{peres1996separability}, generically, it is an NP-hard problem~\cite{gurvits2003classical}. Besides, even though at least one linear entanglement witness could be found to witness any entangled state~\cite{horodecki1996teleportation,terhal2000bell,horodecki2009quantum,guhne2009entanglement} as displayed in Fig.~\ref{Fig:entanglement-witness}, there still lacks a universal and scalable way to construct such an appropriate witness for an arbitrary state in practice.

In this work, we turn to the machine learning technique which is powerful in extracting features or patterns hidden in large multipartite datasets to tackle the quantum detection problem. Recently, much progress has been achieved in this inter-disciplinary field of quantum machine learning~\cite{Biamonte2017}. For example, on one hand, many quantum or quantum-inspired algorithms have been developed to speed up some well-known machine learning algorithms~\cite{xiao2010quantum,lloyd2014quantum,tang2019quantum}. On the other hand, machine learning is also a natural candidate to extract correlated features of high-dimensional quantum systems, which has found wide applications in quantum control~\cite{bukov2018reinforcement}, state tomography~\cite{chapman2016experimental}, measurement~\cite{magesan2015machine,hentschel2010machine}, and many-body problems~\cite{carleo2017solving,huang2017accelerated,carrasquilla2017machine}. Especially, the task of quantum entanglement detection can be formulated as a binary classification problem. As a consequence, various classical neural nets, trained with both entangled and separable samples, have been constructed to solve this problem via supervised learning~\cite{lu2018separability,yang2019experimental,ma2018transforming}. However, the supervised training method requires a large pre-labelled dataset. In practice, it is time-consuming or even impossible to faithfully label a large number of entangled states in a high-dimensional space~\cite{gurvits2003classical}, thus leading these supervised methods into a dilemma.

Here, we instead build up an unsupervised model to accomplish the task of entanglement detection beyond the above issues. Following from the fact that separable states form a convex set, it becomes an anomaly detection problem of which all separable samples are labelled as normal and entangled ones are abnormal. Particularly, as shown in Fig.~\ref{Fig:pipeline}, a class of complex-valued neural networks composed of a pseudo-siamese network and a Generative Adversarial Net (GAN), is constructed and then trained with very few normal samples to detect entanglement for multipartite systems, ranging from 2-qubit to 10-qubit states. It is noted that our model is much more feasible than anomaly detection methods proposed in~\cite{Liu2018,Liang2019} which require quantum hardware.


It is further illustrated in Fig.~\ref{Fig:entanglement-witness} that our unsupervised neural nets are essentially trained to search for proper nonlinear entanglement witnesses which near-perfectly construct the boundary between separable and entangled samples. Numerical results show that it is able to achieve extremely high accuracy of entanglement detection with above $97.5\%$ on average, and even capable to detect partial entanglement within subsystems, e.g., bi-separable states in 3-qubit system with accuracy above $97.7\%$.

Our work is organised as follows. In section~\ref{sec2}, we give a brief introduction to the task of entanglement detection and unsupervised learning method. Then we propose an unsupervised learning neural network targeted for the detection of generic quantum features. In section~\ref{sec3}, multipartite entanglement detection is taken as examples to illustrate the performance of our model, with only separable samples used for training. Finally, we conclude this work with a summary in section~\ref{sec:con}.

\section{Unsupervised entanglement detection}\label{sec2}

\subsection{The task of detecting entanglement}

Entanglement is not only of significant importance to understand quantum theory at the fundamental level~\cite{horodecki2009quantum}, but also has found applications in information protocols, such as quantum teleportation~\cite{bennett1993teleporting}. For a given $n$-partite quantum system, entanglement associated with the state is defined in a passive way in which a state $\rho$ is entangled if and only if it cannot be described in a fully-separable form of~\cite{werner1989high}
\begin{equation}
\rho_{\rm sep} = \sum_{i=1}^m\lambda_{i}\rho^1_i \otimes \cdots \otimes\rho^j_i\otimes \cdots \otimes \rho^n_i
\label{Eq:separable-state-def}
\end{equation}
with non-negative coefficients satisfying $\sum_{i=1}^m\lambda_{i} = 1$. Here $\rho^j_i$ denotes the state density matrix of the $j$-th subsystem. Obviously, all of the separable states as per \Eref{Eq:separable-state-def} form a convex set in the sense that any convex combination of these states in this set also belong to the same state set. It is noted that the above definition of entanglement does not fully capture the entangled structure in the state, e.g., the partial entanglement~\cite{dur1999quantum}, which will be discussed later.

In practice, whether a given state $\rho$ is entangled or not, can be experimental-friendly determined via an entanglement witness~\cite{horodecki2009quantum,guhne2009entanglement}. Indeed, as shown in \Fref{Fig:entanglement-witness}{\bf (a)}, an entanglement witness essentially defines a hyperplane which separates the entangled state from the convex set of separable states. Furthermore, it has been shown in \cite{horodecki2009quantum} that it is impossible for one linear witness to detect all entangled states, implying that a large set of linear witnesses illustrated in \Fref{Fig:entanglement-witness}{\bf (b)} (could be impractical) or certain nonlinear witness shown in \Fref{Fig:entanglement-witness}{\bf (d)} may be required. Besides, it becomes extremely inefficient and impractical to construct a proper witness for an arbitrary state, especially in multipartite systems. The entanglement witnesses as neural networks are experimentally accessible and has been demonstrated in \cite{ma2018transforming}. In fact, since neural networks are learning the linear and nonlinear correlations on the quantum states to form a classifier, a properly parameterized neural network layer is equivalent to a set of generalized Bell's inequalities for the experimental detection of entanglement. In the following, we propose a complex-valued neural network trained in unsupervised manner to search for the nonlinear entanglement witnesses as desired.


\subsection{Unsupervised learning}
The unsupervised model refers to the process of learning a probability
distribution over the data that has not been classified or categorized. In this situation, automated methods or algorithms must explore the underlying features from the available data and group them with similar characteristics. Specifically, the unsupervised model only receives a training set $\cal{S}$ that contains
\begin{equation}
	{\cal S}  = \{x_1, x_2, x_3, \cdots \}
\end{equation}
without supervised target outputs $\{y_1, y_2, y_3, \cdots \}$. In contrast to supervised learning where tagging data requires a large amount of time, unsupervised learning exhibits high efficiency and self-organization in capturing patterns from untagged data.

\begin{figure}[t]
	\includegraphics[scale=1]{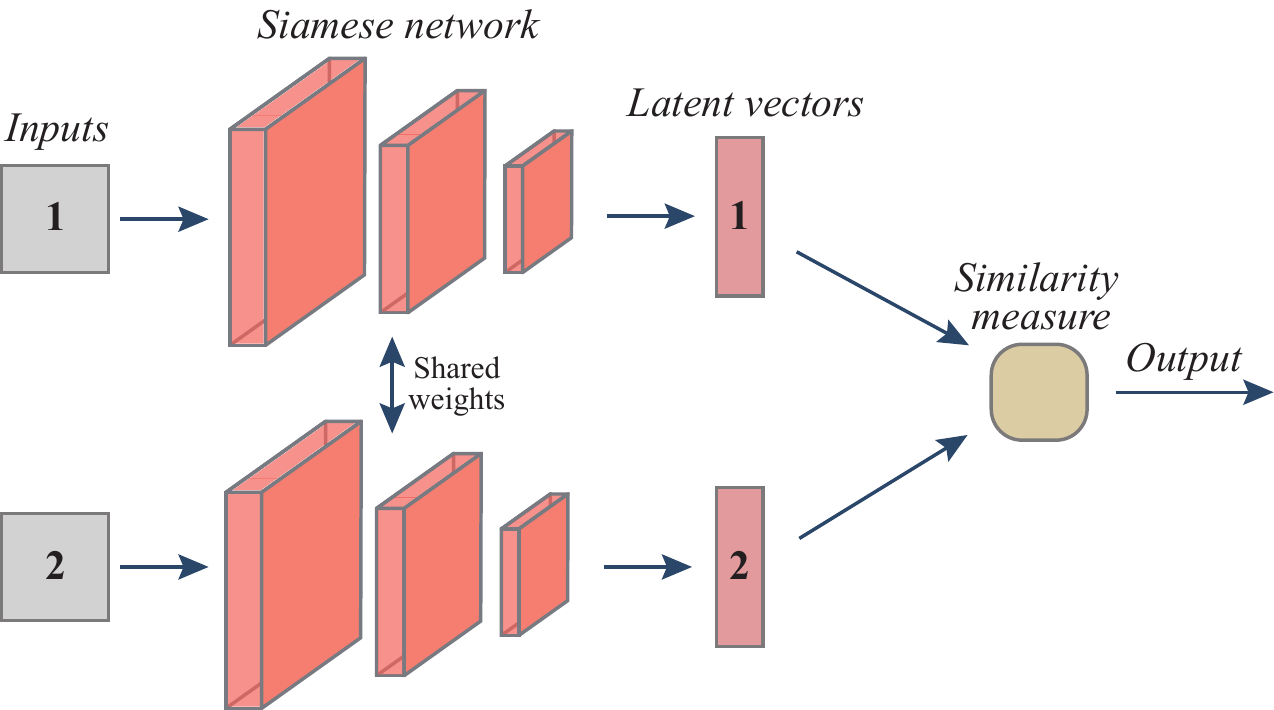}
	\centering
	\caption{
	\label{Fig:siamese}
	\textbf{Pipeline of the siamese network.} Two inputs are encoded into latent vectors whose difference is detected by a similarity measure.}
\end{figure}

Autoencoder \cite{baldi2012autoencoders} is a widely used unsupervised learning method that aims to learn efficient representations for a set of data. Typically, an autoencoder consists of two modules, namely encoder $\mathcal{E}$ and decoder $\mathcal{D}$, where the former learns the latent representation (encoding) for input data, and the latter is trained to generate an output as close as possible to its original input from the latent representation. Another well-known unsupervised learning method is GAN \cite{goodfellow2014generative,arjovsky2017wasserstein}. Specifically, two neural networks, namely generator $\mathcal{G}$ and discriminator $\mathcal{D}$, contest with each other in the form of a zero-sum game in GAN, where the gain of one module is the loss of the other. This technique learns to generate new data with the same statistics as the training set. The siamese network \cite{chicco2021siamese, koch2015siamese}, as shown in \Fref{Fig:siamese}, contains a pair of neural networks built by the same parameters, which receives two inputs and detects their difference by comparing the output vectors of the networks. The siamese network is capable of learning generic features for making predictions about an unknown distribution even when few examples from the distribution are available, which provides a competitive approach for pattern recognition without the domain-specific knowledge. In particular, the siamese network can be trained in an unsupervised manner, as the labels of the input data are not needed.

For these reasons, the method proposed in this paper has been built upon the siamese network, which is suitable for one-class unsupervised learning. The basic idea is similar to one-class support vector machine for anomaly detection \cite{tax2004support}. That is, given a set of training samples, we aim to model the underlying distribution of the data and detect the soft boundary of this set, in order to classify new inputs as belonging to this set or not. In this case, the model will only take a training dataset without class labels as input, which means the model is a type of unsupervised learning methods.

\begin{figure}
	\includegraphics[scale=0.9]{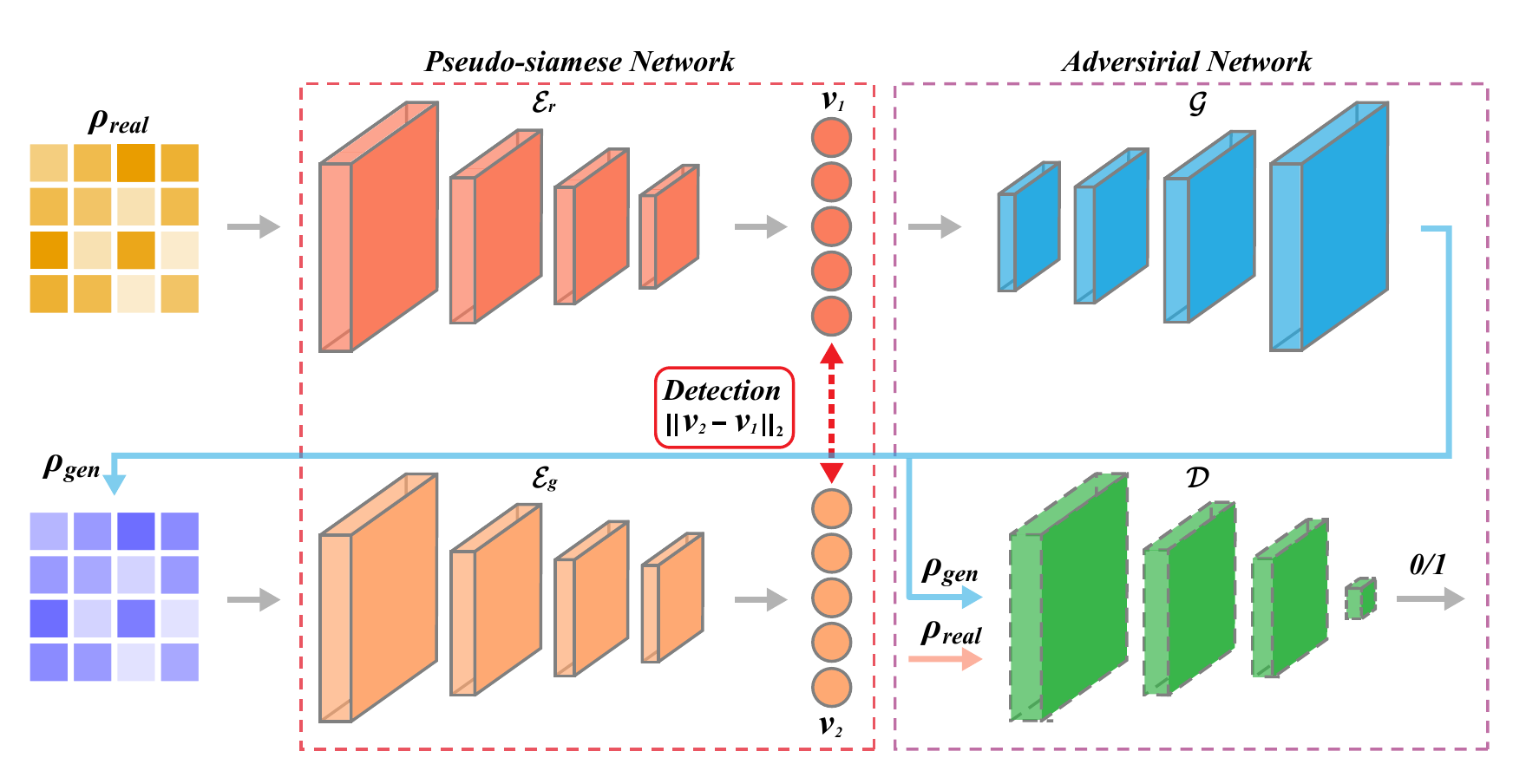}
	\centering
	\caption{
		\label{Fig:pipeline}
		\textbf{Structure of the complex-valued neural network.} The complex-valued network is composed of two parts: One is the pseudo-siamese network in the red dashed box and the other is a GAN in the purple dashed box. The pseudo-siamese neural network consists of two encoders $\mathcal E_r$ and $\mathcal E_g$ that share the same network structure. $\rho_{\rm gen}$ is generated by $\mathcal{G}$. The discriminator network $\mathcal{D}$ is a binary classifier which outputs either $0$ or $1$. The generator $\mathcal{G}$ and discriminator $\mathcal{D}$ form the GAN, which aims to produce a $\rho_{\rm gen}$ that is as close as possible to $\rho_{\rm real}$.
	}
\end{figure}

\subsection{Constructing the complex-valued neural networks}

As shown in \Fref{Fig:pipeline}, our networks could be decomposed into two parts: One is the pseudo-siamese neural network (in the red dashed box) and the other is the GAN (in the purple dashed box). The complex-valued neural network receives the density state matrix as the input. The building modules for these networks are detailed in \ref{append_a}.

The pseudo-siamese neural network consists of two encoders sharing the same network structure, labelled as $\mathcal{E}_r$ and $\mathcal{E}_g$, respectively. In contrast to the original siamese network \cite{koch2015siamese} which requires quadratic pairs as input, the pseudo-siamese network only requires a single input $\rho_{\rm real}$ be fed to the first encoder $\mathcal{E}_r$. The second input $\rho_{\rm gen}$ to the second encoder $\mathcal{E}_g$ is automatically generated by the decoder $\mathcal{G}$ whose aim is to reconstruct $\rho_{\rm real}$. Therefore, the pseudo-siamese network trains much faster than the original siamese network while inherits its few-shot learning ability. In principle, these two encoders competes with each other to produce a pair of indistinguishable feature vectors $\bm v_1$ and $\bm v_2$. The performance is evaluated by the cost function
\begin{eqnarray}
\label{Eq:encoder-cost}
\mathcal{L}_1=\mathbf{E}_{\rho_{\rm real}}\|\mathcal E_r(\rho_{\rm real})-\mathcal E_g(\mathcal G(\mathcal E_r(\rho_{\rm real})))\|=\mathbf{E}_{\rho_{\rm real}}\|\bm v_1-\bm v_2\|,
\end{eqnarray}
where the norm $\|\bm x\|$ could be the $L_p$-norm of any complex vector $\bm x$ with $\|\bm x\|_p\equiv(|\Re(\bm x)|^p+|\Im(\bm x)|^p)^{1/p}$. Here 2-norm is chosen for \Eref{Eq:encoder-cost}. As the two inputs to the encoders $\mathcal{E}_r$ and $\mathcal{E}_g$ are slightly different, the two encoders would not share the same weight parameters after training~\cite{hughes2018identifying}.

Combining the encoder $\mathcal{E}_r$ with $\mathcal{G}$ yields an encoder-decoder structure which aims to produce fake samples that are close to real ones. Thus we introduce the loss function
\begin{equation}
\label{Eq:contextual-cost}
\mathcal{L}_2=\mathbf{E}_{\rho_{\rm real}}\|\rho_{\rm real}-\mathcal G(\mathcal E_r(\rho_{\rm real}))\|=\mathbf{E}_{\rho_{\rm real}}\|\rho_{\rm real}-\rho_{\rm gen}\|
\end{equation}
to quantify its performance. In analogy to classical autoencoders~\cite{isola2017image}, it is found that $L_1$-norm achieves better performance than that of $p=2$ for this loss term.

An optional discriminator network could be introduced for additional adversarial training. The discriminator $\mathcal{D}$ and generator $\mathcal{G}$ form the GAN (Fig.~\ref{Fig:pipeline}) which could enhance the ability of $\mathcal{G}$ to produce more realistic quantum samples. Indeed, $\mathcal{D}$ is a binary classifier trained to discriminate fake samples from real ones. The two cost functions for this adversarial net are given by
\begin{eqnarray}
\mathcal{L}_{\rm adv1}&=\mathbf{E}_{\rho_{\rm real}}(-\mathcal D(\rho_{\rm real})+ \mathcal D\left(\mathcal G(\mathcal E_r(\rho_{\rm real}))\right)),\label{Eq:adversarial-cost1}\\
\mathcal{L}_{\rm adv2}&=\mathbf{E}_{\rho_{\rm real}} (- \mathcal D\left(\mathcal G(\mathcal E_r(\rho_{\rm real}))\right))\label{Eq:adversarial-cost2},
\end{eqnarray}
which are alternatively minimized via gradient descent method. Specifically, the gradients are clipped between $-1$ and $1$, turning the network into a Wasserstein GAN which is easy to train~\cite{arjovsky2017wasserstein}. In each round, the parameters of $\mathcal D$ are updated by minimizing $\mathcal{L}_{\rm adv1}$, while the parameters of $\mathcal G$ and $\mathcal E_r$ are updated by minimizing $\mathcal{L}_{\rm adv2}$.

Finally, by combining \eref{Eq:encoder-cost}-\eref{Eq:adversarial-cost2}, the complex-valued neural network is trained by alternatively minimizing $\mathcal{L}_{\rm adv1}$ and
\begin{eqnarray}
\label{Eq:total-cost}
	\mathcal{L}_3=w_1 \cdot \mathcal{L}_{1} + w_2 \cdot \mathcal{L}_{2} +w_a \cdot \mathcal{L}_{\rm adv2},
\end{eqnarray}
with the weight parameters $w_1$, $w_2$, and $w_a$ being chosen adaptively.

\subsection{Training the networks via unsupervised learning}\label{2-1}

Suppose the complex-valued network is trained with separable states only, an entangled state would result in a feature vector $\bm v_{\rm ent}$ distinct from that of the generated one in the latent space. Indeed, the entire training and prediction process can be divided into three steps as follows.
\begin{itemize}
	\item [1)]
	Preparing separable states as training samples. Following \Eref{Eq:separable-state-def}, each $\rho^j_i$ is generated via $HH^\dagger/(\tr{H H^\dagger})$, where $H$ is a complex-valued matrix whose real and imaginary parts of each entry are sampled from independent Gaussian distributions. It is noted that this sampling method could cover the whole space of separable states~\cite{zyczkowski2001induced}.
	\item [2)]
	Training the neural network on the generated set of separable states by alternatively minimizing $\mathcal{L}_{\rm adv1}$ as per \Eref{Eq:adversarial-cost1} and $\mathcal{L}_3$ as per \Eref{Eq:total-cost} via the gradient descent method.
	\item [3)]
	Determining the decision threshold value $b$ on the test set after training. We choose $b$ to satisfy
	\begin{equation}
	\label{Eq:eer}
	\rm \frac{FN}{TP+FN} = \frac{FP}{FP+TN},
	\end{equation}
	where TP, FP, TN, and FN refer to the number counts of true positive, false positive, true negative, and false negative samples. Here, being positive or negative stands for a separable or entangled sample. Choosing $b$ to satisfy \Eref{Eq:eer} implies that the probabilities of misclassifying entangled and separable states are the same on the test set. Hence, if the score of a quantum state is larger than this $b$, then it will be detected as entangled.
\end{itemize}

For each $\rho$ in the test set, its score for entanglement detection can be defined as
\begin{equation}
\label{similarity-score}
\mathcal{A}(\rho)=\left\|\mathcal{E}_r(\rho)-\mathcal{E}_g(\mathcal{G}(\mathcal{E}_r(\rho)))\right\|_2.
\end{equation}
It could be further expressed in a witness-like form of
\begin{equation}
\mathcal{A}(\rho)=\|(\mathcal{W}_{ \mathcal{E}_g } \mathcal{W}_{ \mathcal{G}}- {\mathcal{I}})\mathcal{W}_{\mathcal{E}_r}\cdot vec(\rho)\|_2  = \|\mathcal{W}\cdot vec(\rho)\|_2,
\label{Eq:score}
\end{equation}
where $\mathcal{W}_{{ \mathcal{E}_f }(\mathcal G)}$ denotes the weight tensor which generates the corresponding linear and nonlinear network transformations. For this reason, the neural network model can be regarded as trying to determine the nonlinear witness $\mathcal{W}$ which approximately characterizes the boundary between separable and entangled states, without relying on samples of entangled states during training.

Alternately, there is another way to implement the model for prediction without the test dataset, making both training and prediction independent of any information of entangled states. This is achieved by determining $b$ as
\begin{equation}
b=\max_{\rho_{\rm {sep}}}\mathcal{A}(\rho).
\label{Eq:b-max}
\end{equation}
Obviously, this approach leads to a higher detection accuracy than using \Eref{Eq:eer}. Since both the training and implementation do not rely on entangled samples, this approach is computationally efficient. More importantly, the major advantage of our unsupervised learning framework lies in its scalability, as generating sufficient entangled states for training becomes impractical for high-dimensional quantum systems.

\section{Numerical results}\label{sec3}
\subsection{Evaluation metrics}
We use two evaluation metrics of binary classification in our experiments. The first metric is the Area Under Curve (AUC) of the Receiver Operating Characteristic (ROC) curve, which is created by plotting the True Positive Rate ($\rm TPR = TP/ (TP + FN)$) against the False Positive Rate ($\rm FPR = FP/ (FP + TN)$) using the similarity score defined in (\ref{similarity-score}) for various values of $b$ \cite{brown2006receiver}. The second metric is Equal Error Rate (EER), which is defined as $\rm FN/(TP+FN)$ when \Eref{Eq:eer} holds \cite{zhou2009statistical}.

\begin{figure}[t]
	\includegraphics[scale=0.9]{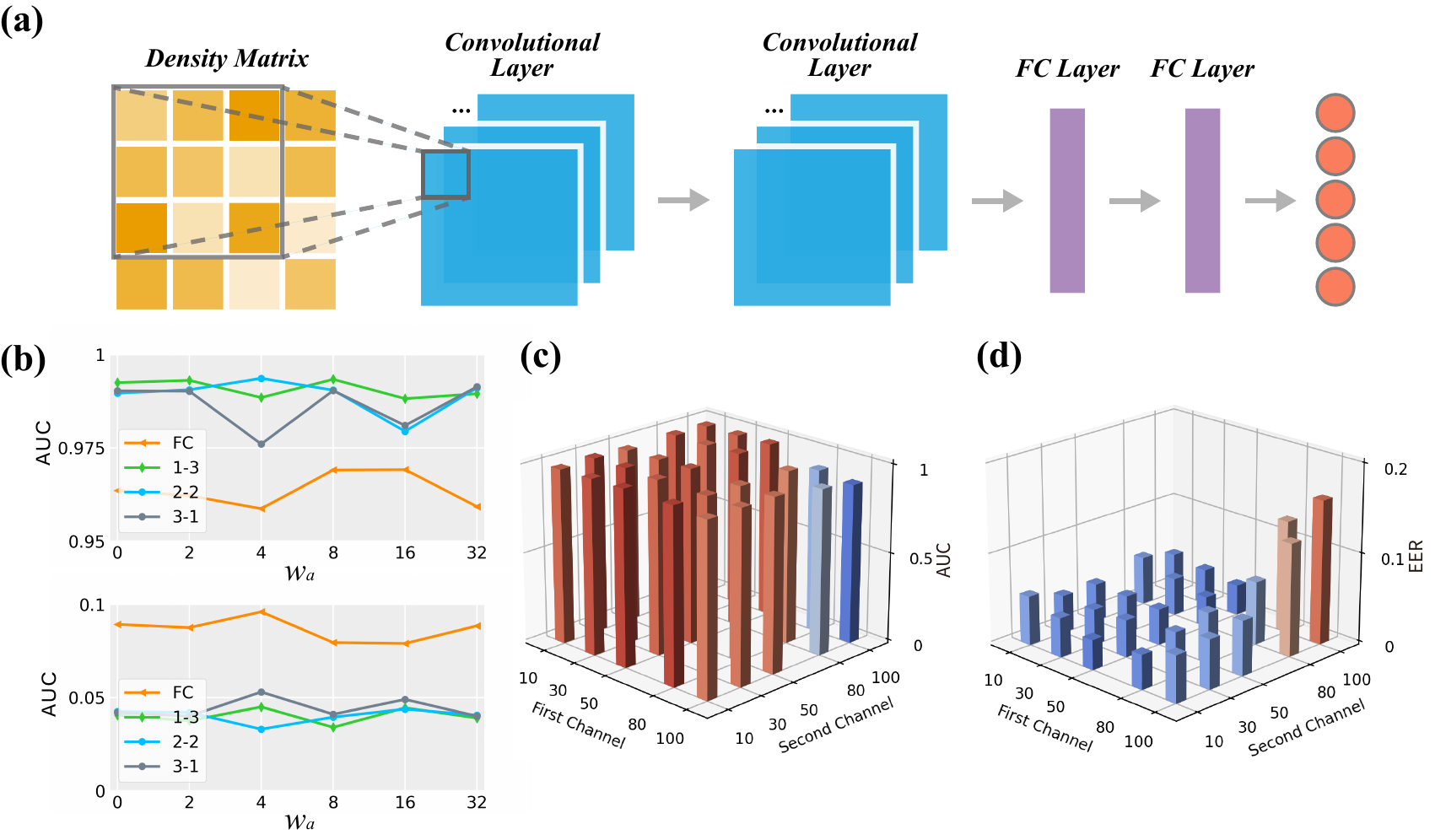}
	\centering
	\caption{
		\label{Fig:result-2qubit}
		\textbf{Detection of 2-qubit states.}
		\textbf{(a)} 4-layer neural network structure of the encoder, including two convolutional layers followed by two FC layers. The structures of the encoder and decoder are symmetric. The structure of the discriminator is the same as the encoder, with an additional normalization layer to produce a scalar output.
		\textbf{(b)} The performance of the neural networks with the first two layers being convolutional or FC. The convolutional kernel size combinations that have been tested are 1 $\times$ 1 and 3 $\times$ 3, 2 $\times$ 2 and 2 $\times$ 2, 3 $\times$ 3 and 1 $\times$ 1, with the number of output channels being $10$ and $30$, respectively. If the first two layers are FC, the number of output channels is set to 32 and 128, respectively.
		\textbf{(e-f)} AUCs and EERs with different number of output channels for convolutional layers. The convolutional kernel sizes are 2 $\times$ 2 and 2 $\times$ 2.
	}
\end{figure}

\subsection{Detecting 2-qubit entangled states}\label{3-1}
The number of training samples for 2-qubit case is 160000, all composed of separable states. The number of testing samples is 80000, including 40000 separable states and 40000 entangled states. 2-qubit separable states are generated by
\begin{equation}
\rho_{\rm {sep}} = \sum_{i = 1}^m\lambda_{i}\rho^1_i \otimes \rho^2_i, \label{Eq:separable-state1}
\end{equation}
where $\sum_{i=1}^m\lambda_{i} = 1$ and $0 \leq  \lambda_{i} \leq 1$, with $m$ iterating from $1$ to $20$. Entangled states are selected from randomly generated states of the entire system using PPT criterion.

The structure of the 4-layer encoder is illustrated in \Fref{Fig:result-2qubit}\textbf{(a)}. The last two layers of the encoder are Fully-Connected (FC) layers, with output channels being 64 and 10, respectively. The first two layers can be convolutional with different kernels and different number of output channels, or fully connected as tested in \Fref{Fig:result-2qubit}\textbf{(b)}. The best performance of the model has been achieved with the convolutional kernel size of the first two layers being 2 $\times$ 2 and 2 $\times$ 2. The best AUC is $0.99$ and EER is $2.99\%$, attained at a small $w_a$ which is the weight of adversarial cost for training. As shown in \Fref{Fig:result-2qubit}\textbf{(c)}, convolutional layer performs much better than FC layer, with AUC being consistently higher than $0.975$ and EER lower than $5\%$. \Fref{Fig:result-2qubit}\textbf{(e-f)} shows the performance of convolutional neural networks when the number of output channels varies, indicating that a small number of output channels is enough to extract the features of entanglement for 2-qubit states.\\

\begin{figure}[t]
	\includegraphics[scale=0.6]{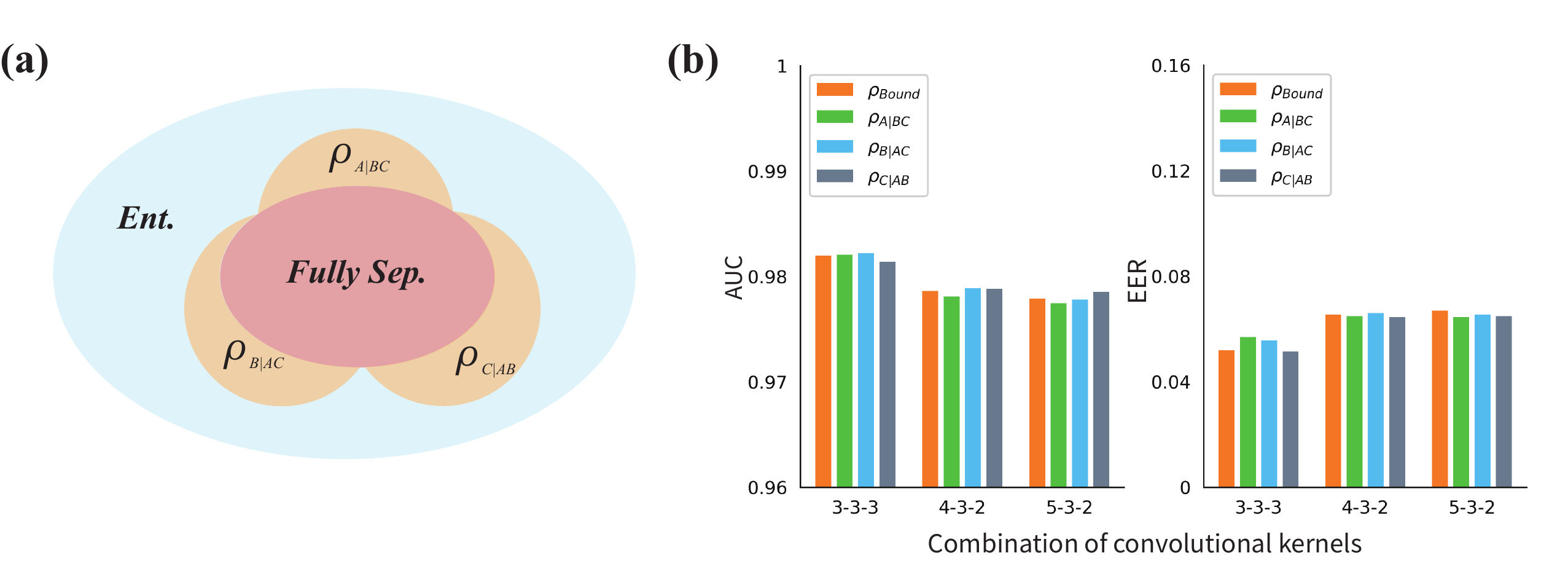}
	\centering
	\caption{
		\label{Fig:result-3qubit}
		\textbf{Detection of 3-qubit states.}
		\textbf{(a)} Distribution of 3-qubit states. $\rho_{A|BC}$ is a bi-separable state with qubit B and C entangled.
		\textbf{(b)} EERs and AUCs with different combinations of convolutional kernels, where $i-j-k$ stands for the kernel sizes of the three convolutional layers.
	}
\end{figure}

\subsection{Detecting 3-qubit entangled states}\label{3-2}

An entangled 3-qubit state can be classified into several types, e.g. bi-separable states and bound entangled states \cite{ma2018transforming}. The 3-qubit state is fully-separable if
\begin{equation}
\rho_{\rm {sep}} = \sum_{i = 1}^m\lambda_{i}\rho^{\rm A}_i \otimes \rho^{\rm B}_i \otimes \rho^{\rm C}_i. \label{separable_state2}
\end{equation}
The distribution of 3-qubit states is illustrated in \Fref{Fig:result-3qubit}\textbf{(a)}. In this case, successful supervised learning requires that one can generate enough and balanced samples for all types of entanglement, which cannot be guaranteed by the current random sampling techniques. In contrast, a universal entanglement detector could be built using only the fully-separable samples if unsupervised learning method is employed.

The numerical results in \Fref{Fig:result-3qubit}\textbf{(b)} are based on a dataset consisting of 160000 training samples and 200000 test samples. The training samples are fully-separable states, and the test samples include 40000 fully-separable states, 40000 bound entangled states and 120000 bi-separable states (40000 for each subtype). To accommodate the $8\times8$ density matrix input, a third convolutional layer is added. The number of the output channels for the three convolution layers is $10,30,50$, respectively. Since the unsupervised model focuses on detecting the feature of separability instead of the features of different types of entanglement, it has achieved similar detection accuracy on four types of entangled samples.

The proposed unsupervised learning method is applicable to the detection of partial entanglement and genuine entanglement. Here we take the detection of bi-separable states of a 3-qubit system as an example \cite{acin2001classification}. Suppose the task is to discriminate the bi-separable states $\rho_{\rm A|BC}$ ($\rm B$ and $\rm C$ are entangled) from the other states. By generating the entangled states for subsystem ${\rm BC}$ using the PPT criterion, the samples of bi-separable states are given by
\begin{equation}
\rho_{\rm A|BC} = \sum_{i = 1}^m\lambda_{i} \rho^{\rm A}_i \otimes \rho^{\rm BC}_i . \label{partial_ent_state}
\end{equation}
A classifier for $\rm A|BC$ separability can be obtained by training on these samples in an unsupervised manner. Particularly, if we replace $\rho^{\rm BC}_i$ in (\ref{partial_ent_state}) by a generic 2-qubit state, the anomalies detected would be the quantum states that are entangled between $\rm A$ and $\rm BC$ (Page 10). Furthermore, if we generate the samples as
\begin{equation}
\rho_{\rm ABC} = \sum_{i = 1}\lambda^1_i \rho^{\rm A}_i \otimes \rho^{\rm BC}_i+\sum_{j = 1}\lambda^2_j\rho^{\rm B}_j \otimes \rho^{\rm AC}_j+\sum_{k = 1}\lambda^3_k \rho^{\rm C}_k \otimes \rho^{\rm AB}_k,\label{genuine_ent_state}
\end{equation}
the abnormal samples detected by the unsupervised model would be quantum states which are not bi-separable. In other words, the genuine entanglement of the 3-qubit state can be detected as an anomaly.

\subsection{Scalability up to 10-qubit states}\label{4}

\begin{figure}
	\includegraphics[scale=0.9]{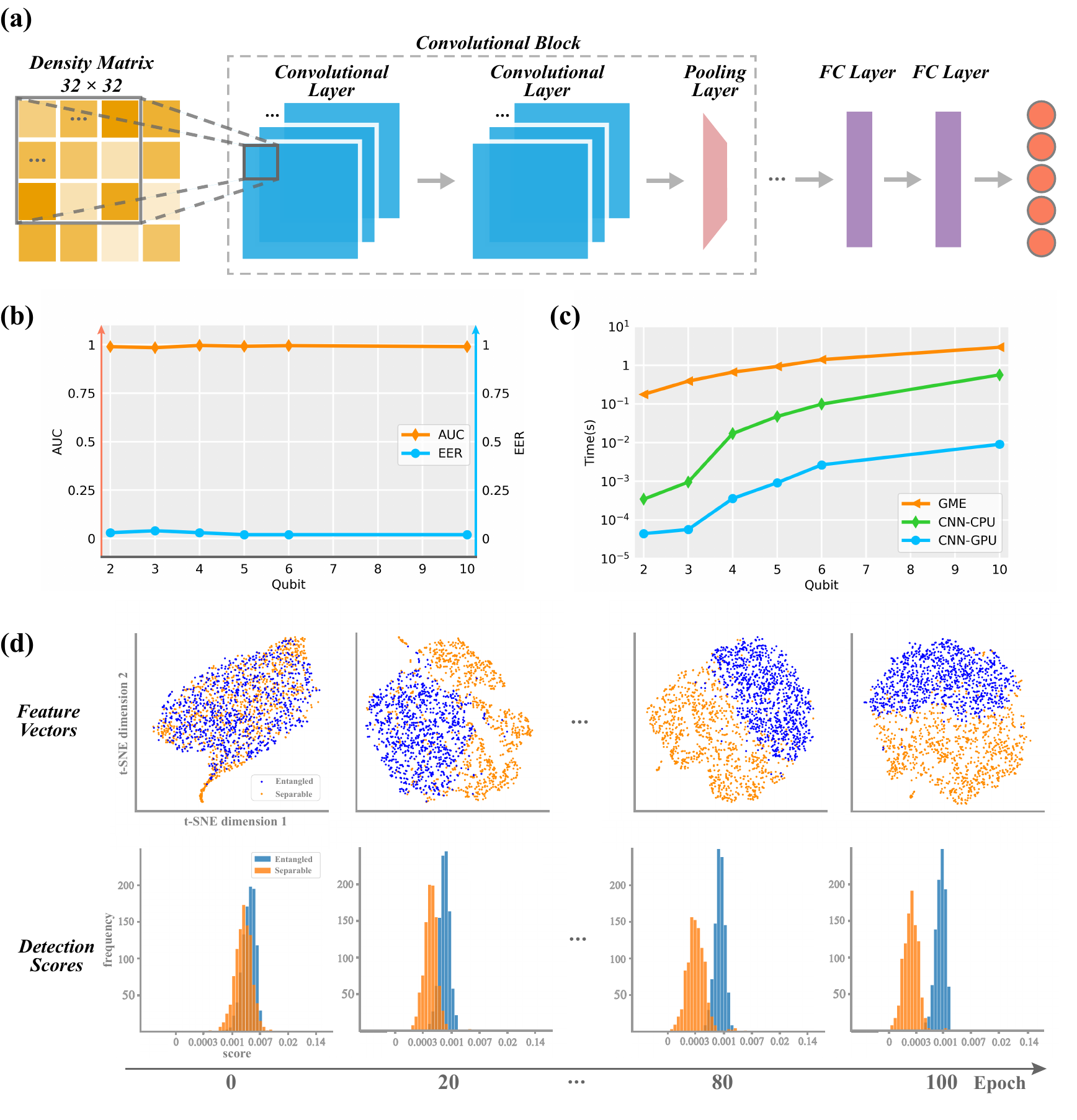}
	\centering
	\caption{
		\label{Fig:result-multiqubit}
		\textbf{Scalability of unsupervised learning.}
		\textbf{(a)} The neural network structure of the encoder for 5-qubit pure states. The convolutional block is composed of two convolutional layers and a $2\times2$ max pooling layer.
		\textbf{(b)} AUCs and EERs achieved by the unsupervised learning method as the number of qubits increases.
		\textbf{(c)} The comparison of inference time between the up-to-date numerical method for computing the Geometrical Entanglement Measure (GME) and the neural network method.
		\textbf{(d)} The evolution of feature vectors and detection scores for 1000 separable and 1000 entangled samples of 5-qubit states during training.
	}
\end{figure}

The unsupervised learning method is applied on 4- to 10-qubit states to study its scalability. We have found that the generation of separable states for training is very efficient even for tens of qubits, because the generation of separable pure states is very efficient, which is done by generating single qubit states and calculating their Kronecker products. Consequently, mixed (fully and partial) separable states can be constructed as linear combinations of pure states, which does not take much time. In this work, it takes less than 10 minutes to generate enough pure separable samples for 10-qubit states on a desktop computer, and mixing the samples takes less than 3 minutes. Moreover, we have observed a linear increase on the generation time with the dimension. Here we used pure 4- to 10-qubit states for training because the test samples of mixed states (mixed entangled states) are hard to label for high-dimensional system, while we have developed an efficient algorithm \cite{zhang2020iterative} that can tell whether a randomly generated 10-qubit pure state is entangled or not within 5 seconds. However, test samples are just used to measure the accuracy of the model. The model is trained using the separable samples only, which can be generated efficiently. The trained model can be implemented without using test samples as shown in (\ref{Eq:b-max}). Therefore, the model can also be trained and implemented with mixed state samples for high-dimensional cases. Note that the geometrical measure is only used to label the entangled states for the test dataset. The separable pure states are generated by
\begin{equation}
\arrowvert\psi_{\rm{sep}}\rangle = \arrowvert\psi^1_i\rangle  \otimes \cdots   \arrowvert\psi^j_i\rangle  \cdots \otimes \arrowvert \psi^n_i\rangle,  \label{separable_state3}
\end{equation}
where $\arrowvert\psi^j_i\rangle$ is a randomly generated pure state vector of the $j$-th qubit. The real and imaginary parts of the complex-valued vector are sampled from an independent Gaussian distribution. The density matrix $\rho_{\rm {sep}} = \arrowvert\psi_{\rm{sep}}\rangle\langle\psi_{\rm{sep}}\arrowvert$ is used as the input to the neural network. \Fref{Fig:result-multiqubit}\textbf{(a)} depicts the network structure of the encoder for entanglement detection in 5-qubit states, where a max pooling layer has been added to handle the increased dimension of the input. For 10-qubit states, we adopt three convolutional layers and increase the max pooling size to $4\times4$. The training dataset is composed of 160000 separable states, and the test dataset is composed of 40000 separable and 40000 entangled states. The entangled states are found by randomly generating 4- to 10-qubit pure states and computing their entanglement measures using the numerical method from \cite{zhang2020iterative}. See \ref{append_b} for the details of the algorithm.

As shown in \Fref{Fig:result-multiqubit}\textbf{(b)}, the unsupervised model achieves an AUC of $0.9952$ and an EER of $2.02\%$ for entanglement detection in 10-qubit states. The EER is $0.54\%$ for entanglement detection in 5-qubit states, which means only 54 in 10000 states are misclassified. The short inference time is another advantage of the neural network model. The inference time of the neural network model on GPU is about tens of microseconds to hundreds of microseconds for up to 10 qubits (\Fref{Fig:result-multiqubit}\textbf{(c)}), which is significantly faster than the up-to-date numerical method which takes the state vector instead of density matrix as the input for computing the geometrical entanglement measure. The time needed for generating training dataset is greatly reduced as compared to supervised learning methods, since there is no need to label the entangled states. For example, suppose the 10-qubit training dataset of the supervised method consists of 100000 samples, which must be labelled by numerically computing the geometrical entanglement measure. The total time needed for generating the dataset is about $138$ hours (labelling each sample takes 5 seconds in average). In contrast, generating separable training samples of the 10-qubit system is much more simple, which only takes several minutes.

The upper half of \Fref{Fig:result-multiqubit}\textbf{(d)} shows the evolution of feature vectors of 1000 separable and 1000 entangled states in the training process for 5-qubit states. We visualize the evolution by t-SNE method \cite{maaten2008visualizing} which maps the feature vectors to two-dimensional space. In the first 10 epochs, the entangled and separable states are mixed up in the latent space and difficult to distinguish. After 20 epochs, the feature vectors start to split into two set. In the last 20 epochs, the feature vectors of separable states are separated completely from the feature vectors of entangled states, with very few exceptions. A similar evolution can be seen in the distribution of detection scores of the input states. After training, the detection scores of separable states are more closed to zero, while the scores of entangled states are concentrated around $0.001$.

\section{Conclusions and Discussions}\label{sec:con}

We have proposed an efficient and scalable method with unsupervised learning to detect quantum entanglement. Specifically, we build up a class of complex-valued pseudo-siamese neural networks which is easy to implement as it is trained without entangled samples. Moreover, it is scalable to detect entanglement of multipartite systems where sufficient labelled entangled samples become difficult to obtain, and our numerical analysis finds that we could still obtain a rather high accuracy with above $97.5\%$ on average for multipartite systems from 2-qubit to 10-qubit. For this reason, we believe that our work provides a promising tool to detect quantum features of high-dimensional quantum data.

Finally, it is noted that we exploit the convexity of separable samples and thus reformulate entanglement detection as an anomaly detection problem, for which the unsupervised neural networks are suitable. Since other useful quantum features, such as Bell nonlocality and Einstein-Podolsky-Rosen steerability, also share the same property that it is defined as a distinguishable sample from a convex set, it is evident that our work can be readily generalized to solve the similar detection problem.

\section*{Acknowledgements}
This research was supported by the National Natural Science Foundation of China under Grants No. 62173296 and No. 62088101. Guofeng Zhang thanks financial support from the Hong Kong Research Grant Council (No. 15208418, No. 15203619, and No. 15506619), Shenzhen Fundamental Research Fund, China, under Grant No. JCYJ20190813165207290, and  the CAS AMSS-polyU Joint Laboratory of Applied Mathematics.

\section*{References}
\bibliographystyle{unsrt}			
\bibliography{ref}

\appendix
\section{Complex-valued neural network}\label{append_a}
We build the complex-valued neural network based on the work of \cite{trabelsi2018deep}. The codes are available at \url{https://github.com/ewellchen/Entanglement_detection}. The two-dimensional convolutional (denoted as $*$) and fully-connected (denoted as $\cdot$) operations of the weight $w$ and input $z$ in the complex domain are defined by
\begin{eqnarray}
w * z &= \Re\{w\} * \Re\{z\} - \Im\{w\} * \Im\{z\} \\
&  +i(\Im\{w\} * \Re\{z\} + \Re\{w\} * \Im\{z\}), \\
w \cdot z &= \Re\{w\}\Re\{z\} -\Im\{w\}\Im\{z\} \\
&+i(\Re\{w\}\Im\{z\}+ \Im\{w\}\Re\{z\}),
\label{CNN and FN}
\end{eqnarray}
where $\Re$ and $\Im$ represent the real and imaginary part of the vector or matrix, respectively. The formulation of the Complex-valued Rectified Linear Unit (CReLU) is given by
\begin{equation}
{\rm CReLU}(z)={\rm ReLU}(\Re\left(z\right)) + i{\rm ReLU}(\Im\left(z\right)),\label{CReLU}
\end{equation}
which introduces nonlinearity into the network transformation. The Batch Normalization (BN) layer is implemented by multiplying the 0-centered data $(\boldsymbol{z}-\mathbf{E}[\boldsymbol{z}])$ with the inverse square root of the covariance as
\begin{eqnarray}	
\mathcal{V}&=&\left(\begin{array}{cc}{\rm{Cov}(\Re\{\boldsymbol{z}\}, \Re\{\boldsymbol{z}\})} & {\rm{Cov}(\Re\{\boldsymbol{z}\}, \Im\{\boldsymbol{z}\})} \\ {\rm{Cov}(\Im\{\boldsymbol{z}\}, \Re\{\boldsymbol{z}\})} & {\rm{Cov}(\Im\{\boldsymbol{z}\}, \Im\{\boldsymbol{z}\})}\end{array}\right),\nonumber\\
\tilde{\boldsymbol{z}}&=&(\mathcal{V})^{-\frac{1}{2}}(\boldsymbol{z}-\mathbf{E}[\boldsymbol{z}]),\nonumber\\
\mathrm{BN}(\tilde{\boldsymbol{z}})&=&\left(\begin{array}{cc}{\gamma_{r r}} & {\gamma_{r i}} \\  {\gamma_{r i}} & {\gamma_{i i}}\end{array}\right) \tilde{\boldsymbol{z}}+\boldsymbol{\beta}.
\label{batch_norm}
\end{eqnarray}
The parameters $\gamma_{r(i)r(i)}$ and $\beta$ are trainable. Each convolutional layer is composed of a convolutional operation, a CReLU and a BN layer. The first fully-connected layer is composed of a fully-connected operation and a CReLU. The last fully-connected layer generates the final output directly via a fully-connected operation. The operations defined above are differentiable, which means the neural network could be trained efficiently with back-propagation. The gradient is calculated with respect to the real-valued cost function $\mathcal L$ as
\begin{equation}
\nabla_{\mathcal L}(z)=\frac{\partial \mathcal L}{\partial z}=\frac{\partial \mathcal L}{\partial z_r}+i\frac{\partial \mathcal L}{\partial z_i}=\Re\left(\nabla_{\mathcal L}(z)\right)+i \Im\left(\nabla_{\mathcal L}(z)\right).\label{gradient}
\end{equation}
The back-propagation updates the complex-valued parameter $t = t_r + i t_i$ of the neural network by
\begin{eqnarray}
\nabla_{\mathcal L}(t) &=\frac{\partial \mathcal L}{\partial t}=\frac{\partial \mathcal L}{\partial t_r}+i \frac{\partial \mathcal L}{\partial t_i} \\
&=\frac{\partial \mathcal L}{\partial z_r} \frac{\partial z_r}{\partial t_r}+\frac{\partial \mathcal L}{\partial z_i} \frac{\partial z_i}{\partial t_r}+i\left(\frac{\partial \mathcal L}{\partial z_r} \frac{\partial z_r}{\partial t_i}+\frac{\partial \mathcal L}{\partial z_i} \frac{\partial z_i}{\partial t_i}\right) \\
&=\frac{\partial \mathcal L}{\partial z_r}\left(\frac{\partial z_r}{\partial t_r}+i \frac{\partial z_r}{\partial t_i}\right)+\frac{\partial \mathcal L}{\partial z_i}\left(\frac{\partial z_i}{\partial t_r}+i \frac{\partial z_i}{\partial t_i}\right) \\
&=\Re\left(\nabla_{\mathcal L}(z)\right)\left(\frac{\partial z_r}{\partial t_r}+i \frac{\partial z_r}{\partial t_i}\right)\\
&+\Im\left(\nabla_{\mathcal L}(z)\right)\left(\frac{\partial z_i}{\partial t_r}+i \frac{\partial z_i}{\partial t_i}\right),
\label{chain_rule}
\end{eqnarray}
which could be implemented using Pytorch \cite{paszke2017automatic}.

\section{Computing the GME of quantum pure states}\label{append_b}
We employ the algorithm proposed in \cite{zhang2020iterative} to compute the GME for an arbitrary quantum pure state. The algorithm is based on a tensor version of the Gauss-Seidel method for computing unitary eigenpairs (U-eigenpairs) of a non-symmetric complex tensor $\mathcal{A}$ which corresponds to the given quantum pure state.

\begin{algo} \cite{zhang2020iterative} \label{algo:1}
	Computing the U-eigenpairs of an $n_1\times\cdots\times n_m$ non-symmetric complex tensor $\mathcal{A}$.
	
	{\bf Step 1 (Initial step)}:
	Let $\mathcal{S}=\mathbf{sym}(\mathcal{A})$ be the symmetric embedding of $\mathcal{A}$, and $n=n_1+\cdots+n_m$. Choose a starting point $\mathbf{x}_0\in\mathbf{C}^n$ with $\|\mathbf{x}_0\|=1$, and $0<\alpha_\mathcal{S}\in \mathbf{R}$. Let $\lambda_0=\mathcal{S}^*\mathbf{x}_0^m$.
	
	{\bf Step 2 (Iterating step)}:
	
	\hskip 0.5 in {\bf for} $k=1,2,\cdots$, {\bf do}
	\begin{eqnarray}
		\label{eqalgorithm1.1} \hat{\mathbf{x}}_k&=&\lambda_{{k-1}}\mathcal{S}\mathbf{x}^{*m-1}_{k-1}+\alpha_\mathcal{S}\mathbf{x}_{k-1},\\
		\label{eqalgorithm1.2} \mathbf{x}_k &=&\hat{\mathbf{x}}_k/\|\hat{\mathbf{x}}_k\|,\\
		\label{eqalgorithm1.3} \lambda_k &=&\mathcal{S}^*\mathbf{x}_k^m.
	\end{eqnarray}
	
	\hskip 0.5 in {\bf end for.}
	
	{\bf return}:
	
	\hskip 0.5 in  Unitary symmetric eigenpair  (US-pair):  $\lambda_\mathcal{S}=|\lambda_k| $, and  $\mathbf{x}=(\frac{\lambda_\mathcal{S}}{\lambda_k})^{1/m}\mathbf{x}_k$.
	
	\hskip 0.5 in Let $\mathbf{x}=(\mathbf{x}^{(1)\top},...,\mathbf{x}^{(m)\top})^\top$, $\mathbf{x}^{(i)}\in \mathbf{C}^{n_i},$ for all $i=1:m$.
	
	\hskip 0.5 in U-eigenvalue $\lambda_\mathcal{A}=\frac{({\sqrt{m}})^m}{m!}\lambda_\mathcal{S}$.
	
	\hskip 0.5 in U-eigenvector $\{\sqrt{m}\mathbf{x}^{(1)},\ \cdots,\ \sqrt{m}\mathbf{x}^{(m)}\}$.
\end{algo}

\end{document}